\documentclass[12pt]{iopart}
\usepackage{iopams}
\usepackage{bm}
\usepackage{epsfig}
\usepackage{graphicx}
\usepackage{amsfonts}
\usepackage{amssymb}
\usepackage{mathrsfs}
\usepackage{color}
\usepackage{appendix}

\begin{document}

\title{All-optical controlled phase gate in quantum dot molecules}
\author{Li-Bo Chen$^{1}$ and Wen Yang$^{2}$}

\begin{abstract}
We propose a two-qubit optically controlled phase gate in quantum dot
molecules via adiabatic passage and hole tunneling. Our proposal combines
the merits of the current generation of vertically stacked self-assembled
InAs quantum dots and adiabatic passage. The simulation shows an
implementation of the gate with a fidelity exceeding 0.98.
\end{abstract}

\pacs{78.67.Hc, 03.67.Lx}

\address{$^{1}$ School of Science, Qingdao Technological University, Qingdao
266033, China}
\address{$^{2}$ Beijing Computational Science Research Center, No.3 Heqing Road Haidian
District Beijing 100084, China}
\eads{\mailto{wenyang@csrc.ac.cn}}
\maketitle
\section{Introduction}

The spin of an electron, trapped in a self-assembled semiconductor quantum
dot (SAQD) and manipulated by laser pulses, is believed to be a promising
qubit candidate for quantum computation and quantum communication. Such
qubits can be optically controlled in high speed \cite{Press,Kim} and
its coherence times has been prolonged to the order of microseconds \cite%
{Greilich}. Recently, there have been many experimental demonstrations of
the key DiVincenzo requirements \cite{DiVincenzo} for such qubit, for
examples, spin initialization \cite{Atature,Emary,X.Xu}, the coherent
manipulation of electron spins \cite{Press,Kim}, and fast spin
nondestructive measurement \cite{Kim2}.

Entangling gates for two qubits form an essential ingredient of quantum
computation. Significant effort has been invested in theoretical protocols
for two qubit gate \cite%
{Piermarocchi,Calarco,Nazir,Gauger,Emary2,kjxu,Puri,LBChen,Economou} and
experimental demonstration of optical entanglement control \cite{D.kim}
between QDs. However there has been no reported experimental realization of
optically controlled phase gate between QDs. In this paper, we propose a
two-qubit controlled phase gate in a vertical self-assembled quantum dot
molecule (SAQDM) \cite{ga1,ga2} utilizing adiabatic passage and hole
tunneling. Compared with our previous scheme \cite{LBChen}, the using of
adiabatic passage efficiently suppressing spontaneous decay from excited
states \cite{kuk,gau,ber,Saikin,Openov,Brandes} and need not precisely
timing the interval between the two pulses. Our proposal bases on the
current generation of SAQDMs, and the simulation shows the gate can
implemented with a fidelity exceeding $0.98$.

\section{The Basic model}

Our model is based on the vertically coupled SAQDs \cite{ga1} (see Fig. 1).
The electrons or holes can tunnel between the two dots to create a quantum
dot molecule (QDM). For the QDMs discussed here, the nominal height of dot $1
$, $h_{1}$, is greater than that of do t $2$, $h_{2}$, so that dot $1$
exhibits the lower transition energy. This allows the hole levels to be
brought into resonance with a positive electric field applied along the $x$
axis (the growth direction), while the electron level of dot $2$ is shifted
to a much higher energy than that of dot $1$. We use the spin of the
electrons in each dot as the qubits. In the Faraday geometry, the energy
levels for the single QD system in the presence of a small magnetic field $%
B_{z}$ along the $z$ axis (perpendicular to the $x$ axis) \cite%
{X.Xu,Berezovsky}. The two eigenstate states of electron spin, i.e. $%
\left\vert \uparrow \right\rangle $ and $\left\vert \downarrow \right\rangle
$, are split by the magnetic field $B_{z}$. The lowest-energy interband
transition is to the trion state $\left\vert \uparrow \downarrow \Uparrow
\right\rangle $ ($\left\vert \downarrow \uparrow \Downarrow \right\rangle $%
), consisting of two electrons in a singlet state and a heavy hole. Optical
selection dictate that the $\sigma ^{-}$ ($\sigma ^{+}$) polarization laser
could coupled the transition $\left\vert \downarrow \right\rangle
\rightarrow \left\vert \downarrow \uparrow \Downarrow \right\rangle $ ($%
\left\vert \uparrow \right\rangle $ $\rightarrow \left\vert \uparrow
\downarrow \Uparrow \right\rangle $) , and other transitions are forbidden.

\section{Implementation of two qubits phase gate}

The ideal phase gate aims to impose a phase change on the state $\left\vert
\uparrow ,\downarrow \right\rangle $ without affecting the phase of the
other three states. It should also preserve phase coherence for a
superposition of the four QD spin states. This operation can be
characterized by the unitary transformation:%
\begin{eqnarray}
\left\vert \uparrow ,\uparrow \right\rangle &\rightarrow &\left\vert
\uparrow ,\uparrow \right\rangle \ ,\left\vert \uparrow ,\downarrow
\right\rangle \rightarrow -\left\vert \uparrow ,\downarrow \right\rangle ,\,
\nonumber \\
\left\vert \downarrow ,\uparrow \right\rangle &\rightarrow &\left\vert
\downarrow ,\uparrow \right\rangle ,\left\vert \downarrow ,\downarrow
\right\rangle \rightarrow \left\vert \downarrow ,\downarrow \right\rangle .
\end{eqnarray}%
We use the convention that the vertical arrows on the left and right of the
comma sign are, respectively, the directions of the spins in dot $1$ and dot
$2$.

Fig. 2 shows the the level diagram of the two quantum dot and gate operation
process, which includes three sequential steps:

Firstly, the QDM is illuminated with a $\sigma ^{-}$ circularly polarized
continuous wave (CW) laser $\Omega _{1}\left( t\right) $ propagating in the $%
x$ direction. The laser is tuned such that it could
create an exciton in the quantum dot $1$ (only if it is state is $\left\vert
\downarrow \right\rangle $) without affecting the quantum dot $2$. The
Hamiltonian for the QDM under this laser excitation is

\begin{equation}
\begin{array}{ccc}
H_{1}= & \Omega _{1}\left( t\right) \left\vert \downarrow \uparrow
\Downarrow ,\downarrow (\uparrow )\right\rangle \left\langle \downarrow
,\downarrow (\uparrow )\right\vert  &  \\
& +\tau \left\vert \downarrow \uparrow ,\Downarrow \downarrow (\uparrow
)\right\rangle \left\langle \downarrow \uparrow \Downarrow ,\downarrow
(\uparrow )\right\vert +H.c.,\  &
\end{array}%
\end{equation}%
where we assume $\hbar =1$, and $\tau $ is the hole tunneling rate between
the two dots. The Hamiltonian has a dark state%
\begin{equation}
\left\vert D\right\rangle \varpropto \tau \left\vert \downarrow ,\downarrow
(\uparrow )\right\rangle -\Omega _{1}\left( t\right) \left\vert \downarrow
\uparrow ,\Downarrow \downarrow (\uparrow )\right\rangle .\,
\end{equation}%
If we turn on the $\sigma ^{-}$ circularly polarized laser and increase the $%
\Omega _{1}\left( t\right) $ slowly, when $\Omega _{1}\left( t\right) \gg
\tau $ the state $\left\vert \downarrow ,\downarrow (\uparrow )\right\rangle
$ will be adiabatically transferred to the state $\left\vert \downarrow
\uparrow ,\Downarrow \downarrow (\uparrow )\right\rangle $ almost without
exciting the media state $\left\vert \downarrow \uparrow \Downarrow
,\downarrow (\uparrow )\right\rangle $. In this process, the states $%
\left\vert \uparrow ,\uparrow \right\rangle $ and $\left\vert \uparrow
,\downarrow \right\rangle $ are not affected by the CW laser $\Omega _{1}$.

Secondly, we apply a $\sigma^{-}$ circularly polarized pulse $\Omega
_{2}$ to coupled the state $\left\vert \uparrow,\downarrow\right\rangle $ to
the state $\left\vert \uparrow,\downarrow\uparrow\Downarrow\right\rangle $.
The Hamiltonian for the QDM under this laser excitation is

\begin{equation}
H_{2}=\Omega _{2}\left( t\right) \left\vert \uparrow ,\downarrow \uparrow
\Downarrow \right\rangle \left\langle \uparrow ,\downarrow \right\vert +\tau
\left\vert \uparrow \Downarrow ,\downarrow \uparrow \right\rangle
\left\langle \uparrow ,\downarrow \uparrow \Downarrow \right\vert +H.c..
\end{equation}%
Starting with the initial state $\left\vert \uparrow ,\downarrow
\right\rangle $ the system evolves at the time $t$ to
\begin{equation}
\begin{array}{ccc}
\left\vert \psi \left( t\right) \right\rangle = & \frac{1}{\tau ^{2}+\Omega
_{2}^{2}}(\tau ^{2}+\Omega _{2}^{2}\cos \left( t\sqrt{\tau ^{2}+\Omega
_{2}^{2}}\right) \left\vert \uparrow ,\downarrow \right\rangle &  \\
& -i\Omega _{2}\sqrt{\tau ^{2}+\Omega _{2}^{2}}\sin \left( t\sqrt{\tau
^{2}+\Omega _{2}^{2}}\right) \left\vert \uparrow ,\downarrow \uparrow
\Downarrow \right\rangle &  \\
& +\left( \tau \Omega _{2}\cos \left( t\sqrt{\tau ^{2}+\Omega _{2}^{2}}%
\right) -\tau \Omega _{2}\right) \left\vert \uparrow \Downarrow ,\downarrow
\uparrow \right\rangle ). &
\end{array}%
\end{equation}%
At $t_{1}\sqrt{\tau ^{2}+\Omega _{2}^{2}}=\pi $, $\left\vert \psi \left(
t_{1}\right) \right\rangle =\frac{\tau ^{2}-\Omega _{2}^{2}}{\tau
^{2}+\Omega _{2}^{2}}\left\vert \uparrow ,\downarrow \right\rangle -\frac{%
2\tau \Omega _{2}}{\tau ^{2}+\Omega _{2}^{2}}\left\vert \uparrow \Downarrow
,\downarrow \uparrow \right\rangle $. In the case $\Omega _{2}\gg \tau $, $%
t_{1}\approx \pi /\Omega _{2}$, $\left\vert \psi \left( t_{1}\right)
\right\rangle \approx -\left\vert \uparrow ,\downarrow \right\rangle $. The
state $\left\vert \uparrow ,\downarrow \right\rangle $ acquiring a $\pi $
phase, the transition $\left\vert \downarrow \uparrow ,\Downarrow \downarrow
\right\rangle \rightarrow \left\vert \downarrow \uparrow ,\Downarrow
\downarrow \uparrow \Downarrow \right\rangle $ is blocked because of the
Pauli exclusion principle.

Finally we slowly turn off the CW laser $\Omega _{1}$ so that the state $%
\left\vert \downarrow \uparrow \Downarrow ,\downarrow (\uparrow
)\right\rangle $ can adiabatically transfer back to the state $\left\vert
\downarrow ,\downarrow (\uparrow )\right\rangle $. The system returns to its
original state and only the state $\left\vert \uparrow ,\downarrow
\right\rangle $ acquires the $-1$ factor, i.e., controlled phase gate.

\section{ Simulation and Conclusion}

\bigskip To simulate the system's dynamics, we employ master equation of
density matrix $\rho$ \cite{wall}

\begin{equation}
\frac{d\rho }{dt}=-i\left[ H_{1}+H_{2},\rho \right] +L\left( \rho \right) ,
\end{equation}%
the superoperator $L$ is given by
\begin{equation}
L\left( \rho \right) =\frac{1}{2}\sum_{i=1}^{2}\left( 2L_{i}\rho
L_{i}^{+}-L_{i}^{+}L_{i}\rho -\rho L_{i}^{+}L_{i}\right) ,
\end{equation}%
where $L_{i}=\sqrt{\gamma _{i}}c_{i}$ describes spontaneous photon decay in
QD $i$.

We choose the laser Rabi frequencies  as $\Omega _{1}=\frac{5\sqrt{\pi }}{2t_{0}}\exp \left[
-\left( 0.05t\right) ^{2}/t_{0}^{2}\right] $, $\Omega _{2}=\frac{2\sqrt{\pi }%
}{t_{0}}\exp \left[ -\left( 2t\right) ^{2}/t_{0}^{2}\right] $, with $t_{0}=1$
ps, $\tau =2$ meV, and $\gamma _{1}=\gamma _{2}=1$ ns$^{-1}$. The dynamics
of matrix elements $\rho _{\left\vert \uparrow ,\downarrow \right\rangle
\left\langle \uparrow ,\downarrow \right\vert }^{00}$, $\rho _{\left\vert
\downarrow ,\uparrow \right\rangle \left\langle \downarrow ,\uparrow
\right\vert }^{00}$, and $\rho _{\left\vert \downarrow ,\downarrow
\right\rangle \left\langle \downarrow ,\downarrow \right\vert }^{00}$, are
shown in Fig. 3, $\rho _{\left\vert \uparrow ,\uparrow \right\rangle
\left\langle \uparrow ,\uparrow \right\vert }^{00}$ doesn't change in this
process. It shows that the time of implement the phase gate is $T_{g}\approx
100$ ps, if initial state $\left\vert \Psi ^{0}\right\rangle =\frac{1}{2}%
\left( \left\vert \uparrow ,\uparrow \right\rangle +\left\vert \uparrow
,\downarrow \right\rangle +\left\vert \downarrow ,\uparrow \right\rangle
+\left\vert \downarrow ,\downarrow \right\rangle \right) $, we calculate the
fidelity of the phase gate $F=0.98$. The fidelity can be improved further if
the lifetime of the state $\left\vert \downarrow \uparrow ,\Downarrow
\uparrow \right\rangle $ is increased. Consider this system in a cavity, the
state $\left\vert \downarrow \uparrow ,\Downarrow \uparrow \right\rangle $
will be off resonant with their cavity modes, $\gamma _{2}$ can be reduced
to 1.25 ns$^{-1}$ \cite{Hennessy}, the fidelity can be as high as $0.998$.

In conclusion, we have proposed an all-optical controlled phase gate which
benefits from current generation of QDM and adiabatic passage. Our
simulation shows the fidelity of the phase gate is exceeding $0.98$ by using
realistic values for all parameters.

\ack
We thanks Prof. L. J. Sham for helpful discussions. This research was
supported by the National Natural Science Foundation of China (Grant
No.11304174), Natural Science Foundation of Shandong Province (Grant No.
ZR2013AQ010), and the research starting foundation of Qingdao.

\begin{figure}[tbph]
\includegraphics[scale=1, angle=0]{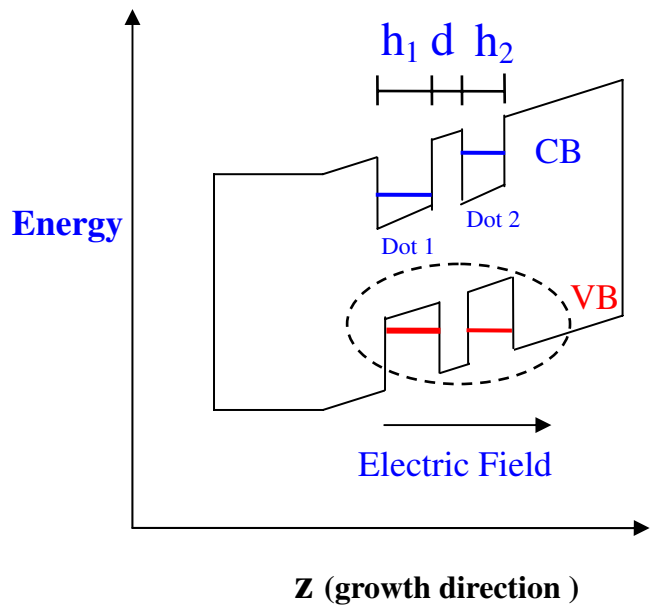}
\caption{Schematic of the vertically coupled quantum dot system. The height
of the dot 1 is $h_{1}$, that of dot 2 is $h_{2}$, the interdot barrier is $%
d $. A positive electric field along the z axis is applied to bring the hole
levels into resonance.}
\end{figure}

\begin{figure}[tbph]
\includegraphics[scale=0.6, angle=0]{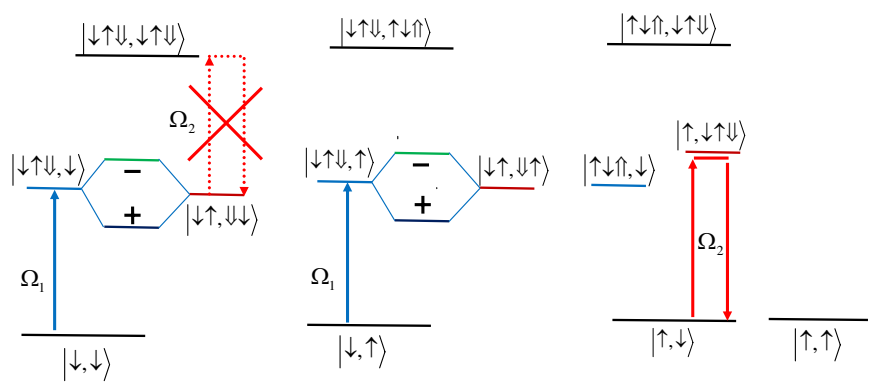}
\caption{The energy diagram of the two Quantum dot ground states and the
optically allowed transitions to the trion states. The large X indicates
that the $\Omega_{2}$ pulse does not affect this transition while performing
a $2\protect\pi$ rotation between $\left\vert
\uparrow,\downarrow\right\rangle $ and $\left\vert
\uparrow,\downarrow\uparrow\Downarrow\right\rangle $.}
\end{figure}

\begin{figure}[tbph]
\includegraphics[scale=1, angle=0]
{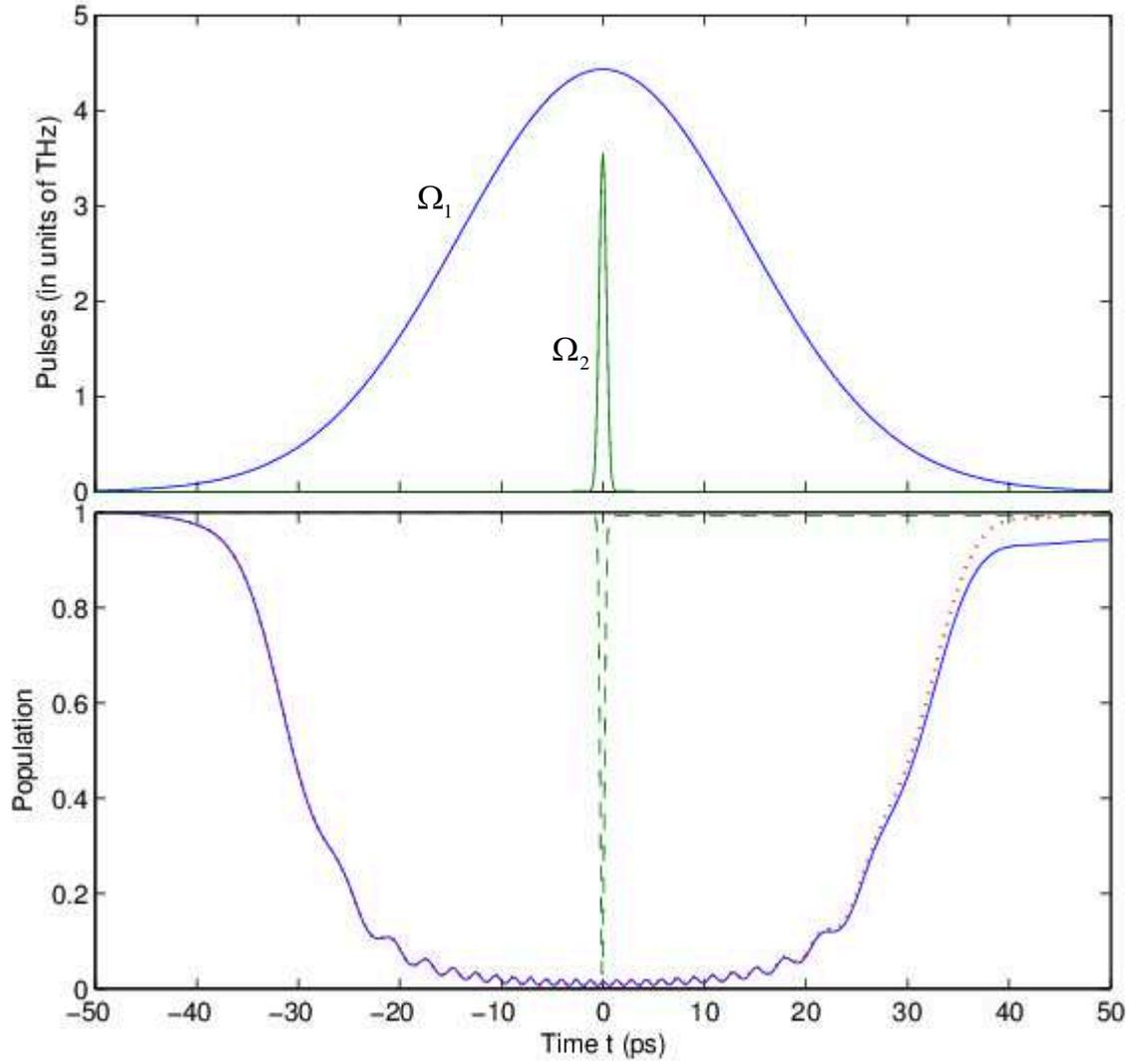}
\caption{Dynamical evolution of several density matrix elements during the
gate operation via numerical simulation. The top figure shows the pulse
intensity of $\Omega_{1}$ and $\Omega_{2}$. The bottom figure shows the time
evolution of three density matrix elements during the controlled phase gate
operation, in which the solid (blue) line, the dot (red) line, and the dash
(green) line denote density matrix elements $\protect\rho_{\left\vert
\downarrow,\uparrow \right\rangle \left\langle
\downarrow,\uparrow\right\vert }^{00}$, $\protect\rho_{\left\vert
\downarrow,\downarrow\right\rangle \left\langle
\downarrow,\downarrow\right\vert }^{00}$, and $\protect\rho_{\left\vert
\uparrow ,\downarrow\right\rangle \left\langle
\uparrow,\downarrow\right\vert }^{00}$. $\protect\rho_{\left\vert
\uparrow,\uparrow\right\rangle \left\langle \uparrow ,\uparrow\right\vert
}^{00}$ doesn't change in this process.}
\end{figure}

\end{document}